\documentclass[twocolumn,aps,prd,preprintnumbers,nofootinbib]{revtex4}
\usepackage[dvipdfmx]{graphicx}
\usepackage{bm,latexsym,amsmath,amssymb,amsfonts,color}

\usepackage[normalem]{ulem}

\begin{document} 

\title{\uline{}Wormhole on DGP brane} 

\author{Yoshimune Tomikawa,${}^1$ Tetsuya Shiromizu,${}^{1,3}$ and Keisuke Izumi${}^2$}

\affiliation{${}^1$Department of Mathematics, Nagoya University, Nagoya 464-8602, Japan}
\affiliation{${}^2$Leung Center for Cosmology and Particle Astrophysics, National Taiwan University, 
Taipei 10617, Taiwan}
\affiliation{${}^3$Kobayashi-Maskawa Institute, Nagoya University, Nagoya 464-8602, Japan}
\begin{abstract}
We analyze a spacetime structure on a new brane configuration constructed recently 
in the Dvali-Gabadadze-Porrati braneworld context. 
The brane, embedded on a five-dimensional bubble of nothing, has a wormhole structure. 
It is an exact solution without any matter fields, and thus the energy conditions for 
matter fields are trivially satisfied. We see that, under the traversability condition, 
the size of the bubble should be larger than $10^{10}$~cm or so. 
\end{abstract}
\maketitle

\section{Introduction} \label{intro}

A wormhole has been often discussed as one of the intriguing objects in general relativity 
\cite{M-T, M-T-Y, visser}(see Ref. \cite{lobo2007} for the recent review).  
However, it is well known that, to support a static wormhole, we need exotic matter 
fields violating the null energy condition \cite{H-V static}(see also Refs. \cite{V-K-D, Ida:1999an}).  
Recently, dynamical wormholes have also been discussed 
\cite{H-V dynamic, Hayward:1998pp, M-H-C, roman, K-S, kim}, because the requirement of the staticity 
for spacetimes is rather strong. One may hope that the existence of dynamical wormhole solutions is 
compatible with the energy conditions. There have been other attempts to construct wormholes; for example, 
considering rotational \cite{teo, kuh}, spherically symmetric with the cosmological constant 
\cite{L-L-O}, or plane-symmetric cases \cite{L-L}. 

In this paper, we discuss the wormhole configuration in a braneworld context(see Refs. \cite{r10, r13} for related works, 
but their setups are completely different from the 
current one). Therein, our Universe is a membrane in a higher- dimensional spacetime. There are 
several models for the braneworld(see Ref. \cite{Koyama} for a review). The simplest one is the 
Randall-Sundrum (RS) model \cite{RS}. It turns out that the wormhole is realized there \cite{Ida}. 
This is regarded as a similar configuration with the Kaluza-Klein 
bubble spacetime found in Ref. \cite{Witten}. We call this RS bubble spacetime. The existence of such a 
configuration means that the Minkowski brane in the anti--de Sitter bulk spacetime may decay into 
the RS bubble spacetime by semiclassical instability. However, for the construction of the RS bubble, 
we need a domain wall at the junction ``point" between two branes. 

The Dvali-Gabadadze-Porrati (DGP) braneworld~\cite{dgp} is another well-motivated model, where 
the induced gravity due to the quantum effects of matter \cite{loop} localized on the brane 
is taken into account. The DGP model was one of the candidates to explain the current acceleration 
of the Universe without introducing the cosmological constant. Now it is known that the 
self-accelerating solution in the DGP model has several serious problems: the ghost 
instability \cite{Izumi:2007pb,Izumi:2008st,Koyama:2005tx,Izumi:2006ca,Liu:2014xja}, 
the incompatible situations with observations \cite{Fang}, etc. Meanwhile, the original configuration, 
so-called normal branch solutions, still can work. However, the vacuum state of the normal branch may 
decay into other states by semiclassical instability. In the recent paper~\cite{I-S}, without 
introducing any domain wall on the brane, the brane configuration with the bubble of nothing has 
been constructed in the DGP braneworld model \cite{dgp}, which may be the state after 
semiclassical decay. We call this the DGP bubble(see Ref. \cite{Gregory2007, Izumi2007} for related work). 

The solution found in Ref.~\cite{I-S} itself is interesting, in addition to the discussion of 
semiclassical decay, because the solution seems to have a wormhole structure without introducing 
any exotic matter fields. The energy conditions trivially hold there because it is the solution 
for the pure gravitational system; i.e., no matter fields are introduced both in the bulk and on 
the brane. The purpose of this paper is to examine the details of the spacetime structure on the 
brane obtained in Ref. \cite{I-S}. 

The spacetime that we will analyze here is dynamical. It is nontrivial to define the wormhole throat 
in dynamical spacetimes because of the spacetime foliation dependence. Here we consult the two typical 
definitions based on the double null~\cite{H-V dynamic, Hayward:1998pp} and time slice 
foliations~\cite{M-H-C} and demonstrate two ways. Although the location of the throat depends on 
foliations, we confirm that the throat exists at least within our considerations. There may be a 
nontrivial time slice, which does not contain the throat, like the critical argument for black 
holes \cite{Wald1991}. Even if not so, it is also very interesting to show the stability of the 
existence of the throat. This is beyond the scope of our current study. 

We also address the traversability shortly. Then, to keep the acceleration and tidal force felt by 
travelers mild like the situation near Earth, we see that the size of the bubble should be 
larger than $\sqrt{(c^2/g_\oplus)|\bf \xi|} \sim 10^{10}$~cm, where $g_\oplus$ is the gravitational 
acceleration on Earth and we took $|\bf \xi| \sim 1$~m, which is the typical size of a human body. 

The remaining part of this paper is organized as follows.  In Sec. \ref{setup}, 
we will briefly describe the setup of the DGP model and the construction of a wormhole spacetime on a 
brane. In Sec. \ref{branewormhole}, we will discuss the spacetime structure on the brane. 
In particular, we will focus on the throat structure and the energy conditions for the effective 
energy-momentum tensor computed from the four-dimensional Einstein tensor. In Sec. \ref{travel}, we address the 
traversability by computing the acceleration and tidal force felt by travelers. Finally, we will give a summary and 
discussion in Sec. \ref{summary}.

\section{Setup} \label{setup}

In this section, we review the DGP bubble spacetime obtained in Ref. \cite{I-S}. 
Therein, there are many brane configurations. For simplicity, 
we consider the single-brane solution, that is, the case without nontrivial junctions of branes.
As pointed out tacitly in Ref. \cite{I-S}, 
the brane geometry has a wormhole structure. We will examine this point in the next section. 

The model is described by the action \cite{I-S,dgp} \footnote{
Exactly, the York-Gibbons-Hawking surface term has to be introduced~\cite{York:1972sj,Gibbons:1976ue}.
} 
\begin{eqnarray}
S &=&  2M^3 \displaystyle \int_{\rm bulk} d^5 x \sqrt{-g} ~R \nonumber\\ 
 &&~~ +2M^3 r_{c}\int_{\rm brane}d^4x \sqrt{-q} ~^{(4)} R(q), 
\end{eqnarray}
where $g_{\mu \nu}$, $q_{\mu \nu}$, $R$, and $^{(4)} R(q)$ 
are the metric of the bulk, the metric of the brane, the Ricci scalar of the five-dimensional bulk, and 
the four-dimensional Ricci scalar on the brane, respectively. 
$M$ is the Planck scale in the five-dimensional spacetime, and $r_c$ is a constant having a length scale. 
We adopt the natural unit in the majority of this paper. 

The bulk spacetime satisfies the vacuum Einstein equation, and the Kaluza-Klein bubble spacetime is a 
solution for the bulk~\cite{Witten}:   
\begin{eqnarray}
ds^2 =f(r)d\chi^2+f^{-1}(r)dr^2+r^2 \gamma_{ab} dx^a dx^b,
\end{eqnarray}
where $f(r) =1-\left(r_0 /r \right)^2$ and $r_0$ is the size of the bubble.
$\gamma_{ab}$ is the metric of the unit three-dimensional de Sitter spacetime:
\begin{eqnarray}
\gamma_{ab} dx^a dx^b =-d\tau^2 +\cosh^2 \tau d\Omega_2^2,
\end{eqnarray}
where $d\Omega_2^2$ is the metric of the unit two-dimensional sphere. 

Now we suppose that the brane is located at 
\begin{eqnarray}
\chi =\bar{\chi}(r) ,
\end{eqnarray}
and then the induced metric of the brane can be written in
\begin{eqnarray}
q =\alpha^{-2}dr^2+r^2 \gamma_{ab} dx^a dx^b, \label{yuudou}
\end{eqnarray}
where
\begin{eqnarray}
\alpha :=\left( \bar{\chi}'^2 f +f^{-1} \right)^{-\frac{1}{2}}
\end{eqnarray}
and $\bar{\chi}':=d\bar{\chi}(r)/dr$. 

From the junction condition, we see that $\bar \chi(r)$ follows 
\begin{eqnarray}
\bar{\chi}'=-\dfrac{r_c}{rf\alpha} (1-\alpha^2). 
\end{eqnarray}
Together with the definition of $\alpha$, it tells us that there are 
two solutions for $\alpha$. Here we take one of them, that is, 
\begin{eqnarray}
\alpha^2 =\dfrac{-(r^2-2r_c^2)+\sqrt{r^4-4r_0^2 r_c^2}}{2r_c^2}. \label{alpha}
\end{eqnarray}
With the additional requirement of $r_0 >r_c$, this case gives us  globally regular 
configurations with the bubble of nothing. Note that $\alpha^2$ 
is positive for 
\begin{eqnarray}
r\geq r_{\ast} := \sqrt{r_0^2 +r_c^2} \label{r>r}
\end{eqnarray}
and $0 \leq \alpha^2 < 1$ ($\alpha^2 \to 1$ as $r \to \infty$). Therefore, the trajectory of 
the brane exists only for $r\geq r_\ast$. In addition, the smooth extension to its own copy 
is possible at $r= r_\ast$(for the detailed discussion of the extension, see Ref.~\cite{I-S}). 
Then, in the final configuration, the brane has two asymptotic regions, both of which approach the Minkowski spacetime.

\section{the wormhole on the brane} \label{branewormhole}

In this section, we examine the induced geometry on the brane. We will 
see that it has indeed a wormhole structure. Since we consider the vacuum brane in the vacuum bulk, 
the energy conditions are trivially satisfied. We also confirm that the effective energy-momentum tensor 
computed from the four-dimensional Einstein tensor does not satisfy them. 

\subsection{The spacetime structure on the brane}

We ask if the spacetime has a wormhole structure. To see this, we need to check two 
points~\cite{H-V dynamic,Hayward:1998pp,M-H-C}: (i) no event horizon and (ii) the existence of a throat. 

To see point (i), we follow the argument of Ref. \cite{I-S}. We introduce the double null 
coordinate $u_\pm$ defined as $du_\pm=d\tau\pm dr/(r\alpha)$, and then the metric becomes 
\begin{eqnarray}
ds^2=-r^2du_+du_-+{\cal {R}}^2d\Omega_2^2,
\end{eqnarray}
where ${\cal {R}}=r \cosh \tau$. Since the null expansion rate $\theta _{\pm}$ can be obtained as 
\begin{eqnarray}
\theta_{\pm} =\dfrac{\partial \ln {\cal {R}}}{\partial u_\pm}= \dfrac{1}{2} (\tanh \tau \pm \alpha ),
\end{eqnarray}
we see that the timelike hypersurface ${\cal H}_-$ with $\theta_+=0$, $\theta_-<0$ exists for 
$\tau <0$ and the timelike hypersurface  ${\cal H}_+$ with $\theta_-=0$, $\theta_+>0$ exists for $\tau>0$. 
For $\tau=0$, both $\theta_\pm$ vanish at $r=r_*$. ${\cal H}_-$ is like an apparent horizon, and 
${\cal H}_+$ is like a cosmological apparent horizon. Moreover, $\theta_+>0$ holds for $\tau>0$, and 
the spacetime is regular everywhere. Thus, we can conclude that there is no event horizon. 
Meanwhile, we have a similar structure to black holes in the region of $\tau<0$; that is, there are 
future trapped surfaces($\theta_\pm <0$). Usually, the presence of the future trapped surface implies 
the appearance of spacetime singularities in the future \cite{HE}. However, 
this is not the case, because the energy conditions for the effective energy-momentum tensor are not 
satisfied in the current case as seen later and then the singularity theorems do not hold on the brane.  

Let us examine point (ii). There are several definitions of throat. Since the induced metric on the 
brane is dynamical, it is not necessary to follow the definition in the old paper \cite{M-T}, 
where a throat is defined just as the minimal surface. Nevertheless, it is interesting to look at this. 
This has been already seen from Fig. 1 of Ref.~\cite{I-S} that the geometry on the $\tau=$const slices 
has a minimum radius at $r=r_{\ast}$. 

One may be interested in the definition proposed by Maeda, Harada, and Carr \cite{M-H-C}, 
where the definition of the throat is generalized to Refs. \cite{H-V dynamic,Hayward:1998pp} 
for spherically symmetric spacetimes. According to them, the throat is a surface with a 
minimum area among trapped spheres which satisfies $\theta_+ \theta_->0$ or a bifurcating 
trapped horizon which satisfies $\theta_+ =\theta_- =0$. We have already seen that 
it holds for inside the region surrounded by ${\cal H}_+$ and ${\cal H}_-$. Moreover, we have known that 
the sphere at $r=r_\ast$, which is a trapped sphere or bifurcating trapped sphere,  has the minimum 
radius on $\tau=$const slices. Thus, we can conclude that the throat is located at $r=r_\ast$. This is 
the same result as that in the old manner. However, as expected, this strongly depends on slices. 

In the current coordinate, the $r=$const trajectory is not static even at $r=\infty$. Therefore, it is 
nice to change the coordinate system. As an example, we introduce a new coordinate $(T,R)$ defined as
\begin{eqnarray}
T=rh(r) \sinh \tau ,~~R=rh(r) \cosh \tau,
\end{eqnarray}
where
\begin{eqnarray}
\ln (h(r))=\displaystyle \int^r \dfrac{1-\alpha}{\alpha r} dr
\end{eqnarray}
and $\displaystyle \lim _{r\to \infty} h(r)=1$.
Then, the metric becomes
\begin{eqnarray}
ds^2 =h^{-2} (-dT^2+dR^2+R^2d\Omega_2^2).
\end{eqnarray}
This has the conformally flat form, and on the asymptotic region $\partial_T$ approaches 
the time direction of the Cartesian coordinate for the Minkowski spacetime. Therefore, this 
form of the metric is one of the natural choices.

First, to find the minimal surface, we look for $R=R_\ast$ that minimizes the area 
$4\pi R^2h^{-2}(R,T)$ on $T=$const slices. 
On the minimal surface,
\begin{eqnarray}
0 &=& \partial_R (Rh^{-1}) \nonumber \\
 & = & h^{-1}\Bigl( 1-(1-\alpha)\frac{R^2}{h^2r^2}\Bigr) \nonumber \\
& = & h^{-1}\Bigl( 1-(1-\alpha)\frac{R^2}{R^2-T^2}\Bigr)
\end{eqnarray}
must be satisfied.
That is, $R_\ast$ satisfies 
\begin{eqnarray}
\Bigl(1-\alpha(R_\ast, T) \Bigr) R_\ast^2 = R_\ast^2-T^2. \label{R*}
\end{eqnarray}
It is difficult to solve this, because we cannot have the explicit expression for the function 
$\alpha(R,T)$. However, moving back to the $(r,\tau)$ coordinate, we can obtain the analytic formula 
of the trajectory $r=r_{\rm min}(\tau)$ satisfying Eq. (\ref{R*}): 
\begin{eqnarray}
r_{\rm min}^2(\tau)=r_c^2(1-\tanh^4 \tau )+r_0^2 (1-\tanh^4 \tau )^{-1}. 
\end{eqnarray}
According to Maeda, Harada, and Carr \cite{M-H-C}, this will be regarded as the location of the throat on $T=$const slices. 
We will first see the slice dependence of the location of the throat and then confirm that it is in the 
trapped region. 

For $r_0 >r_c$, we see that 
\begin{eqnarray}
r_{\rm min}^2(\tau)-r_\ast^2 & = & -r_c^2 \tanh^4 \tau+r_0^2 \frac{\tanh^4 \tau}{1-\tanh^4 \tau} \nonumber \\
& \geq & 
-r_0^2\tanh^4 \tau+r_0^2 \frac{\tanh^4 \tau}{1-\tanh^4 \tau} \nonumber \\
& = & r_0^2 \frac{\tanh^8 \tau}{1-\tanh^4 \tau} \geq 0, 
\end{eqnarray}
with equality if and only if $\tau=0$. As a consequence, the location of the throat determined by 
the minimal surface has the slice dependence. This is not a surprising result. It can be seen, 
in general, from the trace of the extrinsic curvature of two-surface $S$, denoted by $k$. 
When we choose a different time slice, the spacelike normal vector is 
shifted to $\tilde{r}_a=\beta r_a \pm \sqrt{1-\beta^2} t_a$, where $r_a$ is the spacelike 
unit normal vector to $S$ on the hypersurface before shifting, $t_a$ is the timelike unit normal 
vector to $S$ with $t_a r^a =0$, and $\beta$ is a constant. 
We denote the trace of the extrinsic curvature of two-surface $S$ on the shifted hypersurface by 
$\tilde k$. $\tilde k$ is related to $k$ in the following formula:
\begin{eqnarray}
k=h^{ab} \nabla_a r_b =\dfrac{1}{\beta} \tilde{k} \mp \dfrac{\sqrt{1-\beta ^2}}{\beta} h^{ab} K_{ab},
\end{eqnarray}
where $h_{ab}$ is the induced metric of $S$ and $K_{ab}$ is the extrinsic curvature of the spacelike 
hypersurface orthogonal to $t^a$. On the minimal surface, $k=0$ has to be satisfied,
which obviously depends on the choice of slices except for the time-symmetric slice ($K_{ab}=0$). 

Next, we will check that the minimal surface $r=r_{\rm min}(\tau)$ is in the trapped region or at 
the bifurcating trapping horizon. The location of ${\cal H}_\pm$ is determined by 
\begin{eqnarray}
\alpha^2|_{{\cal H}_\pm}=\tanh^2 \tau ,
\end{eqnarray}
and then this gives us the equation of the trajectory for ${\cal H}_\pm$ as 
\begin{eqnarray}
r_{{\cal H}_\pm}^2(\tau)=\dfrac{r_c^2}{\cosh^2 \tau}+r_0^2 \cosh^2 \tau. 
\end{eqnarray}
We find 
\begin{eqnarray}
& & r_{\rm min}^2(\tau)-r_{{\cal H}_\pm}^2(\tau) \nonumber \\
& &~~ =-\frac{\sinh^2 \tau}{1+\tanh^2 \tau} \Bigl[r_0^2-r_c^2
\frac{1+\tanh^2 \tau}{\cosh^4 \tau} \Bigr] \leq 0. 
\end{eqnarray}
Thus, the minimal surfaces on $T=$const($\neq 0$) slices are 
in the trapped region, while that on the $T=0$ slice is at the bifurcating trapping horizon.
Therefore, we have completed the proof that $r=r_{\rm min}(\tau)$ is the throat of this wormhole. 

Finally, we consider the other definition of the throat with respect to the double null foliations 
\cite{H-V dynamic,Hayward:1998pp}. Roughly speaking, the throat is the minimal surface on the 
null surfaces. As we discussed in point (i), we have already seen that ${\cal H}_\pm$ are such surfaces.

\subsection{No exotic}

As stressed before, we do not introduce any matter fields. Thus, no exotic matter fields are 
contained in our model. Usually, we often thought that one needs such exotic matter fields, which 
do not satisfy the energy conditions, to maintain the wormhole. To see what happens on the brane, 
it is nice to define the effective energy-momentum tensor $T^{({\rm eff})}_{\mu\nu}$, computed 
from the four-dimensional Einstein tensor to be 
\begin{eqnarray}
{}^{(4)} G_{\mu\nu} =T^{({\rm eff})}_{\mu\nu}. 
\end{eqnarray}
Now each component of the Einstein tensors is computed to be 
\begin{eqnarray}
& & {}^{(4)}G_{rr} =-\dfrac{3(1-\alpha^2)}{\alpha^2r^2}, \\
& & {}^{(4)} G_{ab} =-(1-\alpha^2 - 2r \alpha \alpha ') \gamma _{ab} ,
\end{eqnarray}
where $\alpha ' :=d\alpha /dr$.
We can write the effective energy-momentum tensor in the orthonormalized coordinate: 
\begin{eqnarray}
T^{({\rm eff})}_{\hat{\mu} \hat{\nu}} =\mathrm{diag} [\rho^{({\rm eff})} ,p_{r}^{({\rm eff})} ,p^{({\rm eff})},p^{({\rm eff})}],
\end{eqnarray}
where 
\begin{eqnarray}
& & \rho^{({\rm eff})} =-p^{({\rm eff})} = \dfrac{1}{r^2} (1-\alpha^2-2r \alpha \alpha'), \\
& & p_{r}^{({\rm eff})} =- \dfrac{3}{r^{2}} (1-\alpha^2). 
\end{eqnarray}
It is easy to see $1-\alpha^2-2r\alpha \alpha' <0$, by using 
\begin{eqnarray}
r \alpha\alpha'=\frac{r^2}{\sqrt{r^4-4r_0^2 r_c^2}} (1-\alpha^2),
\end{eqnarray}
derived from Eq. (\ref{alpha}). Then, we can find $\rho^{({\rm eff})} =-p^{({\rm eff})}<0$; that is, 
the weak ($\rho \geq 0, \rho+p_i \geq 0$) and dominant energy conditions 
($\rho \geq 0, -\rho \leq p_i \leq \rho$) are not satisfied. Moreover, we see 
\begin{eqnarray}
\rho^{({\rm eff})} +p_{r}^{({\rm eff})}=-\frac{2}{r^2}(1-\alpha^2+r\alpha\alpha')<0 ,
\end{eqnarray}
and then the null ($\rho+p_i \geq 0$) and strong energy conditions ($\rho+p_i \geq0, \rho+\sum_i p_i \geq 0$) 
are not satisfied. One may be interested in the positivity of 
$\rho^{({\rm eff})}+p_r^{({\rm eff})}+2p^{({\rm eff})}$, 
which is necessary for the strong energy condition although the other necessary condition is violated.
After short computations, we see that it depends on the values of $r_0$ and $r_c$. 
For $r_0>{\sqrt {3}}r_c$, we see that $\rho^{({\rm eff})}+p_r^{({\rm eff})}+2p^{({\rm eff})}$ is always negative. In the case with 
$r_0 \leq {\sqrt {3}}r_c$, $\rho^{({\rm eff})}+p_r^{({\rm eff})}+2p^{({\rm eff})}$ is positive for 
$r_\ast <r < (2/3^{1/4}){\sqrt {r_0r_c}}$ and negative for $r>(2/3^{1/4}){\sqrt {r_0r_c}}$. 

According to Refs. \cite{I-S, Koyama:2005kd}, we know that 
the gravitational equation on the brane can be  written as 
\begin{eqnarray}
{}^{(4)} G_{\mu \nu} =S_{\mu \nu} -E_{\mu \nu},\label{effEin}
\end{eqnarray}
where
\begin{eqnarray}
S_{\mu \nu} & = & r_c^2 \left\{ \dfrac{2}{3} {}^{(4)} R {}^{(4)} R_{\mu \nu} -{}^{(4)} R_\mu^\alpha 
{}^{(4)} R_{\nu \alpha} \right. \nonumber \\
& & \left. +\dfrac{1}{2} q_{\mu\nu} \left( {}^{(4)} R_{\alpha \beta} {}^{(4)} R^{\alpha \beta}
-\dfrac{1}{2} {}^{(4)} R^{2} \right) \right\}
\end{eqnarray}
and $E_{\mu \nu}$ is the electric part of the Weyl tensor defined by 
$C_{\mu M \nu N}n^M n^N$ with the unit normal vector to the brane $n^M$. 
This is the extended version of the argument for the RS model \cite{SMS} into the DGP model.  

It is nice to look at which part is dominant in the right-hand side of Eq. (\ref{effEin}). Note that 
$E^\mu_\nu$ does not depend on $r_c$:  
\begin{eqnarray}
E_{r}^{r} =\dfrac{3r_0^2}{r^4},\qquad E_b^a =-\dfrac{r_0^2}{r^4} \delta_{b}^{a}. 
\end{eqnarray}
$S^\mu_\nu$ can be computed to be 
\begin{eqnarray}
S_r^r =-3r_c^2 \Bigl( \frac{1-\alpha^2}{r^2} \Bigr)^2 ,
\end{eqnarray}
and
\begin{eqnarray}
S_b^a =r_c^2 \frac{1-\alpha^2}{r^2}\Bigl(\frac{1-\alpha^2}{r^2}+\frac{4\alpha \alpha'}{r}  \Bigr) \delta_b^a. 
\end{eqnarray}
It is easy to see that $E^\mu_\nu$ and $S^\mu_\nu$ are the same order values at 
$r=r_\ast$. However, in the asymptotic region, $S^\mu_\nu$ behaves as 
\begin{eqnarray}
S^r_r \simeq -3r_c^2 \Bigl( \frac{r_0^2}{r^4}\Bigr)^2,
\qquad
S^a_b \simeq 5 r_c^2 \Bigl( \frac{r_0^2}{r^4}\Bigr)^2 \delta^a_b,
\end{eqnarray}
and then $E^\mu_\nu$ is dominant. 
 
We emphasize again that we consider the vacuum solution in the DGP model. 
In this sense, no exotic matter fields are introduced, and thus the energy conditions for 
matter fields are never violated. Rather, the four-dimensional gravitational equation 
on the brane is modified from the Einstein equation. Then, the modification plays effectively 
the role of the exotic source in the Einstein equation. 

\section{Traversability} \label{travel}

In this section, we discuss the traversability of the wormhole on the brane by examining 
acceleration and tidal force for travelers. Let us consider the case that a traveler moves 
across the wormhole along the radial direction while an observer stays at $R=$const.
The orthonormal basis of the traveler's proper reference frame, 
$ \bm{e}_{\hat{0}'}, \bm{e}_{\hat{1}'}$ are related to the observer's one,  
$ \bm{e}_{\hat{0}}=h\partial_T, \bm{e}_{\hat{1}}=h \partial_R $, as  
\begin{eqnarray}
& & \bm{e}_{\hat{0}'} =\gamma \bm{e}_{\hat{0}} \mp \gamma v \bm{e}_{\hat{1}}, \\
& & \bm{e}_{\hat{1}'} =\mp \gamma v \bm{e}_{\hat{0}} +\gamma \bm{e}_{\hat{1}}, 
\end{eqnarray}
where $v$ is the traveler's velocity measured by the observer with the four-velocity of $\bm{e}_{\hat{0}}$ 
at the same position and $\gamma :=(1-v^2)^{-1/2}$. As usual or for simplicity, we consider 
the case where $v=$const.

First, we estimate the acceleration of the traveler. It should not exceed the acceleration 
on Earth $g_{\oplus}$ for the traveler to endure the gravitational force~\cite{M-T}. 
The four-velocity of the traveler is $u^\mu=(\bm{e}_{\hat{0}'})^\mu$, and then a 
short calculation gives us the form of the acceleration: 
\begin{eqnarray}
a^\mu & = & u^\nu \nabla_\nu u^\mu  \nonumber \\
& = & \gamma(-\partial_R h \pm v \partial_Th) (\bm{e}_{{\hat 1}'})^\mu \nonumber \\
& = & -\dfrac{1-\alpha}{r} \gamma \left( \cosh \tau \pm v\sinh \tau \right) (\bm{e}_{{\hat 1}'})^\mu.
\end{eqnarray}
For the traveler to endure the gravitational force, we may need to require 
$|a^i| \lesssim g_{\oplus}$, where $a^i$ is the spatial component of $a^\mu$. 
On the throat of $r=r_{\rm min}(\tau)$, $|a^i|$ becomes
\begin{eqnarray}
|a^i|=\left| \dfrac{\gamma}{r_{\rm min}(\tau) \cosh \tau} \left( 1 \pm v\tanh \tau \right) \right|.
\end{eqnarray}
At $\tau=0$, it becomes 
\begin{eqnarray}
|a^i| \big| _{\tau =0} =\frac{\gamma}{r_\ast}. 
\end{eqnarray}
For $v \ll 1$, we have the following constraint: 
\begin{eqnarray}
r_0^2+r_c^2 \gtrsim \Bigl( \frac{c^2}{g_\oplus} \Bigr)^2, \label{const1}
\end{eqnarray}
where we recovered the speed of light $c$. Then, the concrete value of the constraint on $r_0$ 
becomes 
\begin{eqnarray}
r_0 \gtrsim  \frac{c^2}{g_\oplus}\sim 10^{18}~{\rm cm} \sim 1~{\rm pc}  \label{const-1}
\end{eqnarray}
where we used $r_0>r_c$.

Next, we consider the tidal force acting on the traveler. The proper size of the traveler can 
be written by the displacement vector $\xi^\mu$ satisfying  
\begin{eqnarray}
\xi ^{\hat{0}'} =0,\qquad\mbox{i.e.,}\qquad \xi^\mu u_\mu=0. 
\end{eqnarray}
Then the tidal acceleration $\Delta a^{\mu}$ is defined by 
\begin{eqnarray}
\Delta a^{\hat{\mu}'} =- R^{\hat{\mu}'}_{\ \hat{0}' \hat{\mu}' \hat{0}'} \xi ^{\hat{\mu}'}. \label{deltaa}
\end{eqnarray}
The tidal acceleration $\Delta a^{\mu}$ should be enough small for the traveler to endure, which gives a 
requirement $|\Delta a^i| \lesssim g_\oplus$~\cite{M-T}. Some components of the Riemann tensor are 
written as 
\begin{eqnarray}
R_{\hat 0 \hat 1 \hat 0 \hat 1}& = & h^2 (\partial_T^2-\partial_R^2)\ln h, \\
R_{\hat 0 \hat A \hat 0 \hat B}& = & h^2 \delta_{AB}\Bigl[ 
\partial_T^2 \ln h-\frac{1}{R}\partial_R \ln h +(\partial_R \ln h)^2 \Bigr], \nonumber \\
\\
R_{\hat 1 \hat A \hat 1 \hat B}& = & h^2 \delta_{AB}\Bigl[\partial_R^2 \ln h 
+\frac{1}{R}\partial_R \ln h+(\partial_T \ln h)^2 \Bigr], \nonumber \\
\\
R_{\hat 0 \hat A \hat 1 \hat B}& = & h \delta_{AB} \partial_T \partial_R h,
\end{eqnarray}
where $A$ and $B$ are the indices for the angular coordinate of a two-sphere. 
On the traveler's basis, they are 
\begin{eqnarray}
R_{\hat{0}' \hat{1}' \hat{0}' \hat{1}'} =   R_{\hat 0 \hat 1 \hat 0 \hat 1} 
=\dfrac{\alpha}{r} \dfrac{d\alpha}{dr},
\end{eqnarray}
\begin{eqnarray}
& & R_{\hat{0}' \hat{A}' \hat{0}' \hat{B}'} \nonumber \\
& &~= \gamma^2 \Bigl[R_{\hat 0 \hat A \hat 0 \hat B} \mp v (R_{\hat 0 \hat A \hat 1 \hat B}+R_{\hat 1 \hat A \hat 0 \hat B}
)+v^2R_{\hat 1 \hat A \hat 1 \hat B}  \Bigr]  \nonumber \\
&&~= \Biggl[ -\gamma ^{2} \dfrac{r\alpha \frac{d\alpha}{dr} 
+1-\alpha ^{2}}{r^{2}}   (\cosh \tau \pm v\sinh \tau)^2 \nonumber \\
& &~~~~~~~~\qquad\qquad\qquad\qquad\qquad\quad
+\dfrac{\alpha}{r} \dfrac{d\alpha}{dr}  \Biggr]\delta_{AB}
\end{eqnarray}
and so on. On the throat, we see 
\begin{eqnarray}
R_{\hat{1}' \hat{0}' \hat{1}' \hat{0}'} &=& \dfrac{(1-\tanh ^{4} \tau )^{2}}{r_{0}^{2} -r_{c}^{2} (1-\tanh ^{4} \tau )^{2}},
\end{eqnarray}
\begin{eqnarray}
& & R_{\hat{A}' \hat{0}' \hat{B}' \hat{0}'} \nonumber \\
& &~~ = \Biggl[-\dfrac{2 \gamma^{2} r_{0}^{2} (1-\tanh ^{4} \tau)^{2}}{r_{0}^{4} -r_{c}^{4} (1-\tanh ^{4} \tau)^{4}} (\cosh \tau \pm v \sinh \tau)^2   \nonumber \\
& &~~~~ \qquad\qquad\quad
+\dfrac{(1-\tanh ^{4} \tau)^{2}}{r_{0}^{2} -r_{c}^{2} (1-\tanh ^{4} \tau)^{2}} \Biggr] \delta_{AB}. 
\end{eqnarray}
At $\tau=0$, they become
\begin{eqnarray}
R_{\hat{1}' \hat{0}' \hat{1}' \hat{0}'} & = & \frac{1}{r_0^2-r_c^2}, \nonumber \\
R_{\hat{A}' \hat{0}' \hat{B}' \hat{0}'} & = & \Bigl[- \frac{2\gamma^2r_0^2}{r_0^4-r_c^4}+ \frac{1}{r_0^2-r_c^2}
 \Bigr]\delta_{AB}. 
\end{eqnarray}
Then, for $v \ll 1$, the requirement of $|\Delta a^i| \lesssim g_\oplus$ gives
\begin{eqnarray}
r_0^2-r_c^2 \gtrsim \frac{c^2|{\bf \xi}|}{g_\oplus}, \label{const2}
\end{eqnarray}
and, by the concrete value, we have the constraint on  $r_0$ as 
\begin{eqnarray}
r_0 \gtrsim  \sqrt {\frac{c^2}{g_\oplus} |\bf{\xi}|} \sim 10^{10}\Bigl(\frac{|{\bf \xi}|}{1~{\rm m}} 
\Bigr)^{\frac{1}{2}} ~{\rm cm}. \label{const-2}
\end{eqnarray}
The lower bound is a similar distance between Earth and the Moon.

We have two constraints on $r_0$: Eqs.~(\ref{const-1}) and (\ref{const-2}). The former gives a 
stronger constraint than the latter. However, the acceleration strongly depends on the trajectory 
of the traveler. The tidal force can be estimated to be the product of the curvature and the proper 
size of the traveler, and this rough estimation gives Eq. (\ref{const-2}). Therefore, the latter 
constraint would be independent from the choice of the trajectory and 
may be more meaningful  than the former. 

\section{summary and discussion} \label{summary}

In this paper, we have discussed the spacetime structure on a brane embedded in the five-dimensional 
bubble of nothing spacetime. The solution has been recently constructed in the DGP braneworld 
context \cite{I-S}. This may be realized as a semiclassical decay  of the normal branch solution.
Therein, no matter fields are introduced. The boundary of the bubble in the five-dimensional sense, 
which looks like a singularity for four-dimensional observers, expands with almost the speed of light. 

We have confirmed that, even as no exotic matter fields breaking the energy conditions are introduced, 
the brane has a wormhole throat. This is because the four-dimensional effective equation of gravity 
on the brane is modified from that of general relativity. The modification stems from the 
higher-dimensional effects. Calculating the effective energy-momentum tensor that means the deviation 
from general relativity, we have found that it does not satisfy the energy conditions. 
Namely, the realization of the wormhole structure is supported by the higher-dimensional effects of 
gravity. The higher-dimensional effects can be usually categorized into two: the four-dimensional 
curvature squared part $S^\mu_\nu$ and the electric part of the five-dimensional Weyl tensor $E^\mu_\nu$. 
In the wormhole both are the same order, while in the asymptotic region $E^\mu_\nu$ is dominant. 
There is another way to look at the current situation when one remembers the origin of the DGP model. 
Then one can interpret a part of the higher-dimensional effects as the consequence from 
the quantum effects of localized matter. In this sense, this is consistent with the ordinal thought that 
the quantum effects may support the existence of the wormhole.  

The solution may have a large impact on black hole physics in the DGP braneworld. 
This is an example of a disappearing trapped region without exotic matter fields. 
It means that the existence of an apparent horizon does not guarantee that of a black hole. 
This fact makes the discussion of dynamical black holes in the DGP braneworld comprehensive. It may give us a 
good example which describes some aspects of the evaporation process of black holes
(exactly say, not a black hole); that is, the expanding region may appear after the evaporation of a black hole. 

We discussed the two definitions of a wormhole throat, based on the double 
null~\cite{H-V dynamic, Hayward:1998pp} and time slice foliations~\cite{M-H-C}. We have derived 
the location of the wormhole throat in the different definitions. In the definition based on a 
time slice foliation, it strongly depends on the choice of slices. 

We have discussed the requirement for traversability, too. For travelers to endure the 
gravitational force, the acceleration and tidal force that they feel must be small enough. 
We have seen that the discussion of the acceleration constrains the minimum size of the bubble to be $1~{\rm pc}$, while 
from the constraint on the tidal force the size of the bubble should be larger than $10^{10}$~cm. 
Although the former constraint is much stronger than the latter, the latter may be more meaningful. 
This is because the former strongly depends on the trajectory of the traveler. Even if we take the latter, 
the bubble is too large to nucleate in a quantum process and it is hard to realize a traversable 
wormhole in a natural way. To transverse it, we also need a mechanism to keep the size of the bubble compact. 
Matter fields on the brane may induce such a mechanism naturally because of the attractive gravitational 
force. 

For simplicity, we focused on the simplest configuration among DGP bubble spacetimes. There 
may be a more natural configuration for a specific case. This is left for future study. 

\begin{acknowledgments}
Y.T. expresses his gratitude to H.~Kanno, H.~Awata, and T.~Harada. 
T.S. is supported by Grant-Aid for Scientific Research from Ministry of 
Education, Science, Sports and Culture of Japan (No.~21244033 and No.~25610055). 
K.I. is supported by Taiwan National Science Council under Project No. NSC101-2811-M-002-103.
\end{acknowledgments}

 


\begin{thebibliography}{99}

\bibitem{M-T} 
 M.~S.~Morris and K.~S.~Thorne,
  Am.\ J.\ Phys.\  {\bf 56}, 395 (1988).

\bibitem{M-T-Y} 
  M.~S.~Morris, K.~S.~Thorne, and U.~Yurtsever,
  Phys.\ Rev.\ Lett.\  {\bf 61}, 1446 (1988).

\bibitem{visser}
  M.~Visser,
  {\it Lorentzian Wormhole: From Einstein to Hawking}
  (Springer-Verlag, New York, 1996).

\bibitem{lobo2007} 
  F.~S.~N.~Lobo,
  {\it Classical and Quantum Gravity Research}
  (Nova Science, Hauppauge, NY, 2008), pp. 1-78.

\bibitem{H-V static} 
 D.~Hochberg and M.~Visser,
  Phys.\ Rev.\ D {\bf 56}, 4745 (1997).

\bibitem{Ida:1999an} 
  D.~Ida and S.~A.~Hayward,
  Phys.\ Lett.\ A {\bf 260}, 175 (1999).

\bibitem{V-K-D} 
  M.~Visser, S.~Kar, and N.~Dadhich,
  Phys.\ Rev.\ Lett.\  {\bf 90}, 201102 (2003).

\bibitem{roman} 
  T.~A.~Roman,
  Phys.\ Rev.\ D {\bf 47}, 1370 (1993).

\bibitem{K-S} 
  S.~Kar and D.~Sahdev,
  Phys.\ Rev.\ D {\bf 53}, 722 (1996).

\bibitem{kim} 
  S.-W.~Kim,
  Phys.\ Rev.\ D {\bf 53}, 6889 (1996).

\bibitem{H-V dynamic} 
  D.~Hochberg and M.~Visser,
  Phys.\ Rev.\ Lett.\  {\bf 81}, 746 (1998).

\bibitem{Hayward:1998pp} 
  S.~A.~Hayward,
  Int.\ J.\ Mod.\ Phys.\ D {\bf 08}, 373 (1999).

\bibitem{M-H-C} 
  H.~Maeda, T.~Harada, and B.~J.~Carr,
  Phys.\ Rev.\ D {\bf 79}, 044034 (2009).

\bibitem{teo}  
  E.~Teo,
  Phys.\ Rev.\ D {\bf 58}, 024014 (1998).

\bibitem{kuh} 
  P.~K.~F.~Kuhfittig,
  Phys.\ Rev.\ D {\bf 67}, 064015 (2003).

\bibitem{L-L-O}
  J.~P.~S.~Lemos, F.~S.~N.~Lobo, and S.~Quinet de Oliveira,
  Phys.\ Rev.\ D {\bf 68}, 064004 (2003).

\bibitem{L-L} 
  J.~P.~S.~Lemos and F.~S.~N.~Lobo,
  Phys.\ Rev.\ D {\bf 69}, 104007 (2004).


\bibitem{r10} 
  M.~G.~Richarte,
  Phys.\ Rev.\ D {\bf 82}, 044021 (2010).

\bibitem{r13} 
 M.~G.~Richarte,
  Phys.\ Rev.\ D {\bf 87}, 067503 (2013).

\bibitem{Koyama} 
  R.~Maartens and K.~Koyama,
  Living Rev.\ Relativity\  {\bf 13}, 5 (2010).

\bibitem{RS} 
  L.~Randall and R.~Sundrum,
  Phys.\ Rev.\ Lett.\  {\bf 83}, 3370 (1999);
  {\bf 83}, 4690 (1999).

\bibitem{Ida} 
  D.~Ida, T.~Shiromizu, and H.~Ochiai,
  Phys.\ Rev.\ D {\bf 65}, 023504 (2002);
H.~Ochiai, D.~Ida, and T.~Shiromizu,
  Prog.\ Theor.\ Phys.\  {\bf 107}, 703 (2002).

\bibitem{Witten}
 E.~Witten,
  Nucl.\ Phys.\ B {\bf 195}, 481 (1982).

\bibitem{dgp}
  G.~R.~Dvali, G.~Gabadadze, and M.~Porrati,
  Phys.\ Lett.\ B {\bf 485}, 208 (2000).

\bibitem{loop} 
  D.~M.~Capper,
  Nuovo Cimento Soc.\ Ital.\ Fis.\ A {\bf 25}, 29 (1975);
  S.~L.~Adler,
  Phys.\ Rev.\ Lett.\  {\bf 44}, 1567 (1980);
  A.~Zee,
  Phys.\ Rev.\ Lett.\  {\bf 48}, 295 (1982).

\bibitem{Izumi:2007pb} 
  K.~Izumi and T.~Tanaka,
  Prog.\ Theor.\ Phys.\  {\bf 121}, 419 (2009).

\bibitem{Izumi:2008st} 
  K.~Izumi and T.~Tanaka,
  Prog.\ Theor.\ Phys.\  {\bf 121}, 427 (2009).

\bibitem{Koyama:2005tx} 
  K.~Koyama,
  Phys.\ Rev.\ D {\bf 72}, 123511 (2005).

\bibitem{Izumi:2006ca} 
  K.~Izumi, K.~Koyama, and T.~Tanaka,
  J.\ High Energy Phys.\ 04 (2007) 053.

\bibitem{Liu:2014xja} 
  Y.~W.~Liu, K.~Izumi, M.~Bouhmadi-Lopez, and P.~Chen,
  arXiv:1405.0850.

\bibitem{Fang} 
  W.~Fang, S.~Wang, W.~Hu, Z.~Haiman, L.~Hui, and M.~May,
  Phys.\ Rev.\ D {\bf 78}, 103509 (2008).
 

\bibitem{I-S} 
 K.~Izumi and T.~Shiromizu,
  Phys.\ Rev.\ D {\bf 90}, 046005 (2014).

\bibitem{Gregory2007}
 R.~Gregory, N.~Kaloper, R.~C.~Myers, and A.~Padilla,
  J.\ High Energy Phys.\ 10 (2007) 069.

\bibitem{Izumi2007} 
  K.~Izumi, K.~Koyama, O.~Pujolas, and T.~Tanaka,
  Phys.\ Rev.\ D {\bf 76}, 104041 (2007).


\bibitem{Wald1991} 
  R.~M.~Wald and V.~Iyer,
  Phys.\ Rev.\ D {\bf 44}, 3719 (1991).


\bibitem{York:1972sj}
  J.~W.~York, Jr.,
  Phys.\ Rev.\ Lett.\  {\bf 28}, 1082 (1972).

\bibitem{Gibbons:1976ue}
  G.~W.~Gibbons and S.~W.~Hawking,
  Phys.\ Rev.\ D {\bf 15}, 2752 (1977).

\bibitem{HE}
  S.~W.~Hawking and G.~F.~R.~Ellis, {\it Large Scale Structure of Space-Time} 
  (Cambridge University Press, Cambridge, England, 1975). 

\bibitem{Koyama:2005kd} 
  K.~Koyama and R.~Maartens,
  J.\ Cosmol.\ Astropart.\ Phys.\ 01 (2006) 016.

\bibitem{SMS} 
  T.~Shiromizu, K.-i.~Maeda, and M.~Sasaki,
  Phys.\ Rev.\ D {\bf 62}, 024012 (2000).

\end{thebibliography}
\end{document}